\documentclass[fleqn,usenatbib]{mnras}
\usepackage{newtxtext,newtxmath}
\usepackage[T1]{fontenc}
\usepackage{ae,aecompl}

\usepackage{graphicx}	
\usepackage{amsmath}	
\usepackage{amssymb}	
\usepackage{color}

\title[Windy accretion disc sizes]{Winds Can ``Blow Up'' AGN Accretion Disc Sizes}
\author[M. Y. Sun et al.]{
Mouyuan Sun,$^{1, 2}$\thanks{E-mail: ericsun@ustc.edu.cn (MYS)}
Yongquan Xue,$^{1, 2}$\thanks{E-mail: xuey@ustc.edu.cn (YQX)}
Jonathan R. Trump$^{3}$
and Wei-Min Gu$^{4}$
\\
$^{1}$CAS Key Laboratory for Research in Galaxies and Cosmology, 
Department of Astronomy, University of Science and Technology of China, Hefei 
230026, China\\
$^{2}$School of Astronomy and Space Science, University of Science 
and Technology of China, Hefei 230026, China\\
$^{3}$Department of Physics, University of Connecticut, Storrs, CT 06269, USA\\
$^{4}$Department of Astronomy, Xiamen University, Xiamen, Fujian 361005, China
}

\date{Accepted XXX. Received YYY; in original form ZZZ}

\pubyear{2018}

\begin{document}
\label{firstpage}
\pagerange{\pageref{firstpage}--\pageref{lastpage}}
\maketitle

\begin{abstract}
Recent multi-band variability studies have revealed that active galactic 
nucleus (AGN) accretion disc sizes are generally larger than the predictions 
of the classical thin disc by a factor of $2\sim 3$. This hints at some missing 
key ingredient in the classical thin disc theory: here, we propose an accretion 
disc wind. For a given bolometric luminosity, in the outer part of an accretion 
disc, the effective temperature in the wind case is higher than that in the 
no-wind one; meanwhile, the radial temperature profile of the wind case is 
shallower than the no-wind one. In presence of winds, for a given band, blackbody 
emission from large radii can contribute more to the observed luminosity than 
the no-wind case. Therefore, the disc sizes of the wind case can be larger 
than those of the no-wind case. We demonstrate that a model with the accretion 
rate scaling as $\dot{M}_0 (R/R_{\mathrm{S}})^{\beta}$ (i.e., the accretion 
rate declines with decreasing radius due to winds) can match both the inter-band 
time lags and the spectral energy distribution of NGC 5548. Our model can 
also explain the inter-band time lags of other sources. Therefore, our model 
can help decipher current and future continuum reverberation mapping 
observations. 
\end{abstract}



\section{Introduction}
\label{sect:intro}
The cross correlation between two light curves of active galactic nucleus (AGN) 
continuum emission can provide model-independent constraints on the accretion 
disc sizes. The inter-band correlations might be caused by X-ray 
\citep[e.g.,][]{Krolik1991} or far-ultraviolet (FUV) reprocessing 
\citep{Gardner2017}. According to the reprocessing scenario, X-ray or FUV 
emission acts as an external energy source (in addition to the internal viscous 
dissipation) and heats the outer accretion disc. As a result, emission at longer 
wavelengths (e.g., UV, optical and infrared (IR)) varies in response to X-ray 
or FUV variations (i.e., the ``driving light curves'') after a light-travel time 
delay (i.e., continuum reverberation mapping). Therefore, the inter-band time lags 
can be used to test the accretion disc theory. However, the observed inter-band 
time lags appear to be incompatible with the classical thin disc theory of \cite{SSD}. 
Indeed, the observed accretion disc sizes are $2\sim 3$ times larger than the 
theoretical expectations \citep[e.g.,][]{Edelson2015, Fausnaugh2016, Edelson2017, 
Jiang2017, Mudd2017, Starkey2017, Kokubo2018, McHardy2018}. Accretion disc 
sizes inferred from microlensing observations of quasars are also larger than expected 
\citep[e.g.,][]{Morgan2010}. These observational results indicate that some key 
ingredient is missing in the classical thin disc theory. 

Several scenarios have been proposed to explain the larger-than-expected accretion 
disc sizes. For instance, \cite{Dexter2011} suggested that an inhomogeneous accretion 
disc with temperature fluctuations can explain the microlensing results \citep[for 
alternative explanations, see][]{Abolmasov2012, Li2018}. With some further modifications, 
this model can also explain the timescale-dependent color variations \citep{Cai2016, 
Zhu2018}. However, this model cannot explain inter-band time lags 
unless a speculative common large-scale temperature fluctuation is assumed to 
simultaneously operate over every part of the accretion disc \citep{Cai2018}. 
\cite{Gardner2017} argued that the observed inter-band time lags are not the 
simple light-travel timescales but correspond to other physical timescales (e.g., 
the dynamical or thermal timescales). It is also speculated that the 
larger-than-expected time lags are caused by contribution of diffuse continuum 
emission from broad emission line (BEL) region \citep{Cackett2018, McHardy2018}. 
Last but not least, a non-blackbody disc due to electron scattering in the disc 
atmosphere can be used to explain the observed time lags; however, such a disc 
emits too much soft X-ray emission \citep{Hall2018}. 

Here, we propose an alternative model to explain the larger-than-expected accretion 
disc sizes: an accretion disc suffers from significant winds. Winds, 
which can be probed by blueshifted absorption line features \citep[or blueshifted 
emission lines; see, e.g.,][]{Richards2011, Sulentic2017, Sun2018c} in X-ray 
\citep[i.e., ultra-fast outflows, warm absorbers; see][and references therein]{Tombesi2013}, 
UV and optical bands \citep[e.g.,][]{Weymann1991, Murray1995, Trump2006, Filiz2014, 
Grier2015}, are presumably common in AGNs and might be responsible for the scaling 
relations between supermassive black holes (SMBHs) and their host galaxies \citep[i.e., 
AGN feedback; e.g.,][]{Fabian2012, King2015}. To emit the same 
bolometric luminosity ($L_{\mathrm{bol}}$), the effective temperature of 
the outer disc is higher with significant winds than without except for the innermost part 
(see Fig.~\ref{fig:teff}). In the presence of winds, the radial temperature profile is shallower 
than the classical $T(R)\propto R^{-3/4}$ law. As a result, for a given band, blackbody 
emission from large radii can contribute significantly to the total luminosity. 
Therefore, the accretion disc sizes inferred from inter-band time lags are larger 
than those of the classical disc model. 

This work is formatted as follows. In Section~\ref{sect:model}, we explain our 
model in detail and explore the inter-band time lags of NGC 5548. We discuss 
our results in Section~\ref{sect:dis}. We adopt a flat $\Lambda$CDM cosmology 
with $h_0=0.7$ and $\Omega_{\rm{M}} = 0.3$. 

\section{Thin disc With Winds}
\label{sect:model}
\subsection{General arguments}
We consider that the thin disc suffers from significant winds with a radius-dependent 
mass accretion rate, i.e., $\dot{M}$ is a function of radius ($R$). Several 
mechanisms, including line-driving, radiation pressure, and hydrodynamic acceleration, 
have been proposed to accelerate accretion-disc winds. The observed 
winds are unlikely to be driven by one universal mechanism; instead, 
different mechanisms might work on different spatial scales and different ionization states 
\citep[for a review, see, e.g.,][]{Proga2007}. For instance, the line-driving mechanism 
is likely to be efficient on scales of $\gtrsim 300\ R_{\mathrm{S}}$ ($R_S= 2 
GM_{\mathrm{BH}}/c^2$, where $G$, $M_{\rm BH}$ and $c$ are the gravitational constant, 
black hole mass and speed of light, respectively) and for the gas in a 
low-ionization state \citep[e.g., beyond some "shielding" gas that blocks X-ray photons; see, 
e.g.,][]{Murray1995, Proga2004, Higginbottom2014}. We are interested in winds that are 
launched from the inner most stable circular orbit (ISCO) to a 
few thousand $R_S$. On such spatial scales, the line-driven mechanism is likely to be 
inefficient since gas at these small radii is unlikely to be shielded. Instead, magnetohydrodynamic 
acceleration is a promising alternative mechanism \citep[e.g.,][]{Blandford1982, Cao2013, 
Fukumura2014}.

The exact form of $\dot{M} (R)$ 
is not well known for a thin disc with wind. For instance, magnetohydrodynamic 
acceleration depends on the configuration of the unknown magnetic field. 
Magnetohydrodynamic simulations can answer this question in a self-consistent manner; 
however, the radial dynamical range of such simulation is limited 
\citep[e.g.,][]{Proga2003, Ohsuga2009}. Observations have revealed that winds 
exist in different spatial scales, velocities and ionization states and can be explained by winds 
with self-similar density slopes \citep[e.g.,][]{Steenbrugge2005, Behar2009, Tombesi2013}. Motivated 
by these observations, some magnetically driven wind models \citep[e.g.,][]{Blandford1982, 
Fukumura2014}, and some classical analytic works \citep[e.g.,][]{BB1999, Knigge1999, 
Begelman2012} and numerical simulations of hot accretion flows \citep[e.g.,][]{Yuan2012a}, 
$\dot{M}$ can be assumed to be a simple self-similar form, 

\begin{equation}
\label{eq:mdot}
\dot{M} = \dot{M}_0 r^{\beta} \ ,
\end{equation}
where the dimensionless parameter $\beta$ denotes the strength of winds (without 
winds, $\beta=0$); $r=R/R_S$ is the radial distance ($R$) to the central SMBH  in 
units of the Schwarzschild radius; and $\dot{M}_0$ is a normalization factor (i.e., $\dot{M}_0 
r_{\mathrm{in}}^{\beta}$ is the accretion rate at ISCO, 
$r_{\mathrm{in}}$). The corresponding mass outflowing rate of wind is 
\begin{equation}
\label{eq:mout}
\dot{M}_{\mathrm{out}} = \dot{M}_0 r^{\beta} -  \dot{M}_0 r_{\mathrm{in}}^{\beta}\ .
\end{equation}
That is, we assume that the wind disappears at the ISCO, a reasonable 
boundary condition since the ISCO defines the radius at which gas cannot escape. 

\begin{figure}
\includegraphics[width=\columnwidth]{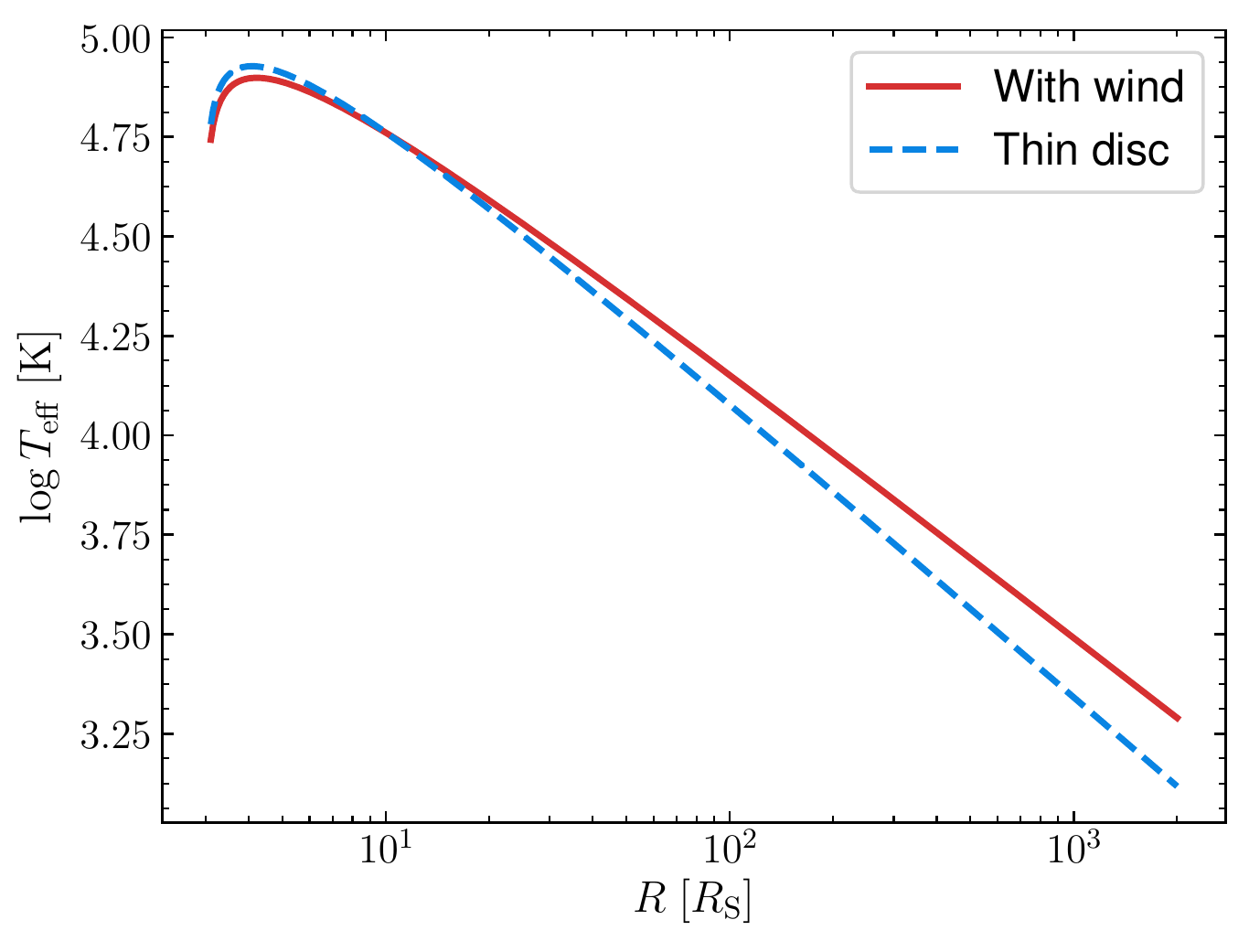}
\caption{The radial distribution of effective temperature ($T_{\mathrm{eff}}$) for cases with the wind 
parameter $\beta=0.3$ (solid line) and $\beta=0$ (i.e., no wind; dashed line), respectively. The 
two discs emit the same bolometric luminosity (i.e., $L_{\mathrm{bol}}=0.1 
L_{\mathrm{Edd}}$); $M_{\mathrm{BH}}$ is $8\times 10^7\ M_{\odot}$. 
At outer radii (i.e., $r>10$), the effective temperature $T_{\mathrm{eff}}$ for the 
wind case is larger than that for the no-wind case.}
\label{fig:teff}
\end{figure}

The disc emits multi-temperature black-body emission. The gravitational 
energy converted into the observed bolometric luminosity is 
\begin{equation}
\label{eq:lbol}
L_{\mathrm{bol}} = \int^{r_{\mathrm{in}}}_{r_{\mathrm{out}}} -\frac{GM_{\mathrm{BH}}\dot{M}}{2 
(rR_S)^2} d(rR_S) = \frac{\dot{M}_0 c^2}{4(1-\beta)r^{1-\beta}_{\mathrm{in}}} \ ,
\end{equation}
where $r_{\mathrm{in}}=3$ for a non-spinning Schwarzschild black hole, and $r_{\mathrm{out}}$ 
($\gg r_{\mathrm{in}}$) is the outer boundary of the accretion disc assumed 
to be the self-gravity radius, beyond which the self gravity dominates over 
the gravity of SMBH. For a SMBH with $M_{\mathrm{BH}}=10^8\ M_{\odot}$, the self-gravity 
radius is approximately $1.8\times 10^3 (\alpha/0.1)^{0.5}\ 
R_S$, where $\alpha$ is the viscosity parameter \citep[see E.q. (4.48) of][]{Netzer2013}. 
For simplicity, we ignore relativistic effects because such corrections are unimportant for 
the UV-optical-IR emission regions. By solving the Navier-Stokes equations 
\citep[e.g.,][]{Bath1983, Knigge1999, Khajenabi2008, Laor2014}, the local effective 
temperature of the disc is
\begin{equation}
\label{eq:teff}
T_{\mathrm{eff}} = \bigg\{\frac{3GM_{\mathrm{BH}}\dot{M}_0 (1- \sqrt{3r_{\mathrm{in}}/r})}{8\pi 
\sigma R_S^3}\bigg\}^{\frac{1}{4}} r^{\frac{\beta-3}{4}} \ ,
\end{equation}
where $\sigma$ is the Stefan-Boltzmann constant. For simplicity, we ignore the 
radiative transfer of the wind; please refer to Section~\ref{sect:rt} for a discussion of the possible 
radiative transfer effects. In Fig.~\ref{fig:teff}, we show the temperature 
profile for the no-wind and the wind with $\beta=0.3$ cases. $\dot{M}_0$ is chosen such 
that the two cases have the same $L_{\mathrm{bol}}$. The temperature profile of the wind case is 
shallower than that of the no-wind case, which is consistent with some microlensing 
studies \citep[e.g.,][]{Poindexter2008, Bate2018}.\footnote{\cite{Bate2018} pointed 
out that previous microlensing studies that found steeper temperature profiles are likely to be 
biased (see their section~7.6).} Note that the flatter temperature profile is also 
proposed in accreting white dwarf systems \citep{Rutten1992, Orosz2003} that might also due to 
winds. In addition, in the outer disc ($r>10$), the effective temperature of the wind case 
is also higher than that of the no-wind case. 

\begin{figure}
\includegraphics[width=\columnwidth]{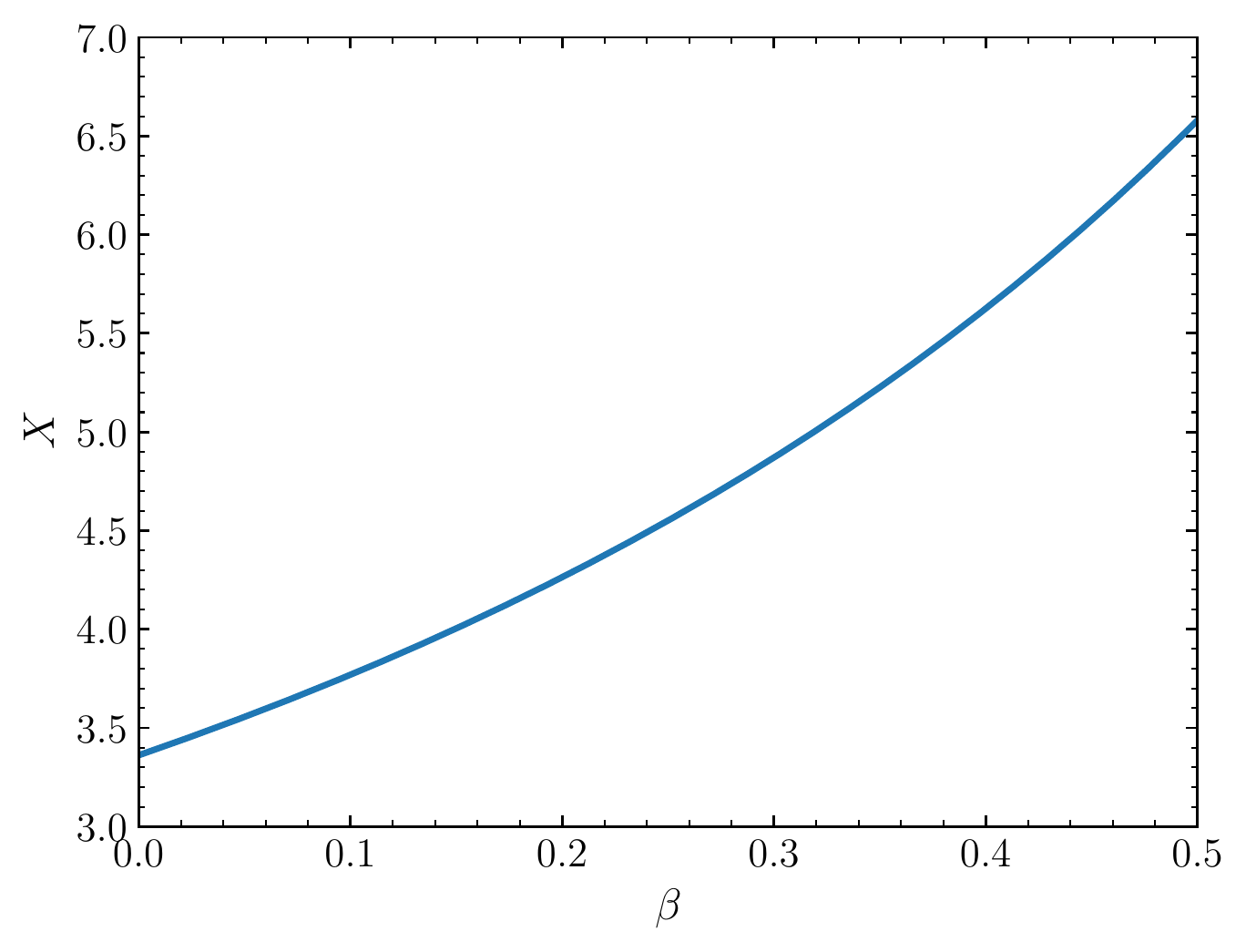}
\caption{The ratio of the flux-weighted flux radius to the peak temperature radius 
$X=R_{\mathrm{fl}}/R_{\lambda}$ as a function of the wind parameter $\beta$ (Eq.~\ref{eq:x}).}
\label{fig:x}
\end{figure}

For a given wavelength $\lambda$, we can define a characteristic radius by setting 
$k_B T_{\mathrm{eff}}=hc/\lambda$, that is, 
\begin{equation}
\label{eq:rlam}
R({\lambda}) = \bigg\{\big(\frac{k_{\mathrm{B}}}{hc}\big)^4 \frac{3G M_{\mathrm{BH}}\dot{M}_0 
(1- \sqrt{3r_{\mathrm{in}}/r})}{8\pi \sigma R_S^{\beta}}\bigg\}^{\frac{1}{3-\beta}} \lambda^{\frac{4}{3-\beta}} \ ,
\end{equation}
where $h$ and $k_{\mathrm{B}}$ are the Planck constant and the Boltzmann constant, 
respectively. In reverberation mapping observations, the measured sizes are not 
$R({\lambda})$ but close to the flux-weighted radius \citep{Fausnaugh2016} if the response 
function is proportional to flux.\footnote{In fact, the measured sizes correspond to the 
response-function-weighted radii, which are generally larger than the flux-weighted radii 
\citep{Hall2018}.} For the face-on case, 
\begin{equation}
\label{eq:rfl}
R_{\mathrm{fl}}({\lambda}) = \frac{\int_{r_{\mathrm{in}}}^{r_{\mathrm{out}}} \pi B(\lambda) 
(rR_S)^2 dr}{\int^{r_{\mathrm{out}}}_{r_{\mathrm{in}}} \pi B(\lambda) rR_S dr} \ ,
\end{equation}
where $B(\lambda)$ is the Planck function. 
If $h\nu/k_{\mathrm{B}}T_{\mathrm{eff}}(r_\mathrm{out})\gg 1$ and $h\nu/k_{\mathrm{B}} 
T_{\mathrm{eff}}(r_\mathrm{in})\ll 1$ (e.g., the optical bands), the ratio $X= 
R_{\mathrm{fl}}/R_{\lambda}$ can be analytically determined. By setting 
$y=hc/(\lambda k_{\mathrm{B}}T)$ and combining Eqs.~\ref{eq:rlam} and \ref{eq:rfl}, 
we find that 
\begin{equation}
\label{eq:x}
X \cong  \frac{\int_{0}^{\infty} \frac{y^{(9+\beta)/(3-\beta)}}{\exp(y)-1} dy}{\int_{0}^{\infty} 
\frac{y^{(5+\beta)/(3-\beta)}}{\exp(y)-1} dy} = \frac{\Gamma(\frac{12}{3-\beta}) 
\zeta(\frac{12}{3-\beta})}{\Gamma(\frac{8}{3-\beta}) \zeta(\frac{8}{3-\beta})} \ ,
\end{equation}
where $\zeta(x)$ and $\Gamma(x)$ denote the Riemann-zeta function 
\citep[][pg. 807]{Abramowitz1970} and the Gamma function, 
respectively; otherwise (e.g., for the near-IR bands), Eq.~\ref{eq:x} overestimates $X$. It is 
clear that $X$ increases with $\beta$ (Fig.~\ref{fig:x}).\footnote{The disc sizes also change with 
inclination angle \citep[see Fig.~2 of][]{Cackett2007}.} The results from Figs.~\ref{fig:teff} and 
\ref{fig:x} suggest that the size of emission region for a given band increases with $\beta$. 

\subsection{The slope of the time lag-wavelength relation}
\label{sect:slope}
For a temperature profile of $T_{\mathrm{eff}}\propto \lambda^{-\alpha}$, the 
time lag-wavelength relation should be $\tau \propto \lambda^{1/\alpha}$ if the light curve is 
infinitely long and the accretion disc has a rather large outer boundary (e.g., $10^5\ R_{S}$). 
That is, the slope of the time lag-wavelength relation for our windy disc is expected to be 
$4/(3-\beta)$ (see Eqs.~\ref{eq:teff} and \ref{eq:rlam}). However, the observed slope can be 
different from $4/(3-\beta)$ 
because of a few reasons. First, the outer boundary sets an upper limit (i.e., 
$R_{\mathrm{out}}/c$) for the inter-band time lag. If the IR emission regions are close to the 
outer boundary, the time lag of IR emission with respect to UV emission increases 
slower than the $\tau \propto \lambda^{4/(3-\beta)}$ relation. Second, the lag estimate is 
often subject to a large variance because of finite-duration monitoring and long-timescale 
trends \citep{Welsh1999}. Third, the induced temperature variations might not be very small. 
Under such circumstances, the temperature profile can be altered by the temperature variability. 
Note that a not-so-small temperature variation does not necessarily conflict with the observed small 
variability amplitudes of UV-Optical-IR emission. This is because the observed UV-Optical-IR 
emission is an integration of blackbody radiation of multiple emission regions; the integration 
process suppresses the variability. Without reprocessing simulations to account for such variance, 
it is inappropriate to infer temperature profile from the time lag-wavelength relation derived 
with short ($\sim 10^2$ days) light curves. Indeed, as we will 
show in Section~\ref{sect:5548} and Fig.~\ref{fig:tlag}, the slope varies significantly per each 
simulation and the slope of the median of the time lag-wavelength relation is actually 
shallower than $4/(3-\beta)$.

\subsection{NGC 5548 as a test case}
\label{sect:5548}
To compare our model with the observed inter-band time lags of NGC 5548, we model a thin 
disc with winds; the physical parameters are $M_{\mathrm{BH}}=8\times 10^7\ M_{\odot}$ 
\citep[see][]{Bentz2010, Pei2017} and the redshift $z=0.017175$ 
\citep{DeRosa2015}. The inclination angle is unknown; we choose a representative 
value of $i=\pi/4$. The SMBH is assumed to be a non-spinning Schwarzschild 
black hole. The inner and outer boundaries of the thin disc are $3\ R_{\mathrm{S}}$ 
and $2000\ R_{\mathrm{S}}$, respectively. As for the wind parameter $\beta$, we test 
two cases, $\beta=0.3$ and $\beta=0.5$. $\dot{M}_0$ is selected in such a way that 
$L_{\mathrm{bol}}=l_{\mathrm{Edd}} L_{\mathrm{Edd}}$, where $L_{\mathrm{Edd}}=1.26\times 
10^{38}M_{\mathrm{BH}}/M_{\odot}\ \mathrm{erg\ s^{-1}}$. We also test two cases for 
$l_{\mathrm{Edd}}$, $l_{\mathrm{Edd}}=0.05$ and $l_{\mathrm{Edd}}=0.1$. Therefore, 
we explore four cases: $l_{\mathrm{Edd}}=0.05$ and $\beta=0.3$; $l_{\mathrm{Edd}}=0.05$ and 
$\beta=0.5$; $l_{\mathrm{Edd}}=0.1$ and $\beta=0.3$; $l_{\mathrm{Edd}}=0.1$ and $\beta=0.5$.

This disc is illuminated with a driving light curve; the emission region of the driving 
light curve is assumed to be a point source located above the SMBH with a scale height of 
$H=5\ R_{\mathrm{S}}$. It is argued that the observed X-ray light curves are not consistent 
with the driving light curve \citep[][but see \citealt{Sun2018b}]{Gardner2017, Starkey2017}. 
Therefore, we assume the driving light curve and the $1367$ \AA\ light curve are 
similar. The external heating due to the driving light curve illumination at each radius is 
assumed to be $1/3$ of the internal viscous dissipation rate \citep{Fausnaugh2016}. 

\begin{figure}
\includegraphics[width=\columnwidth]{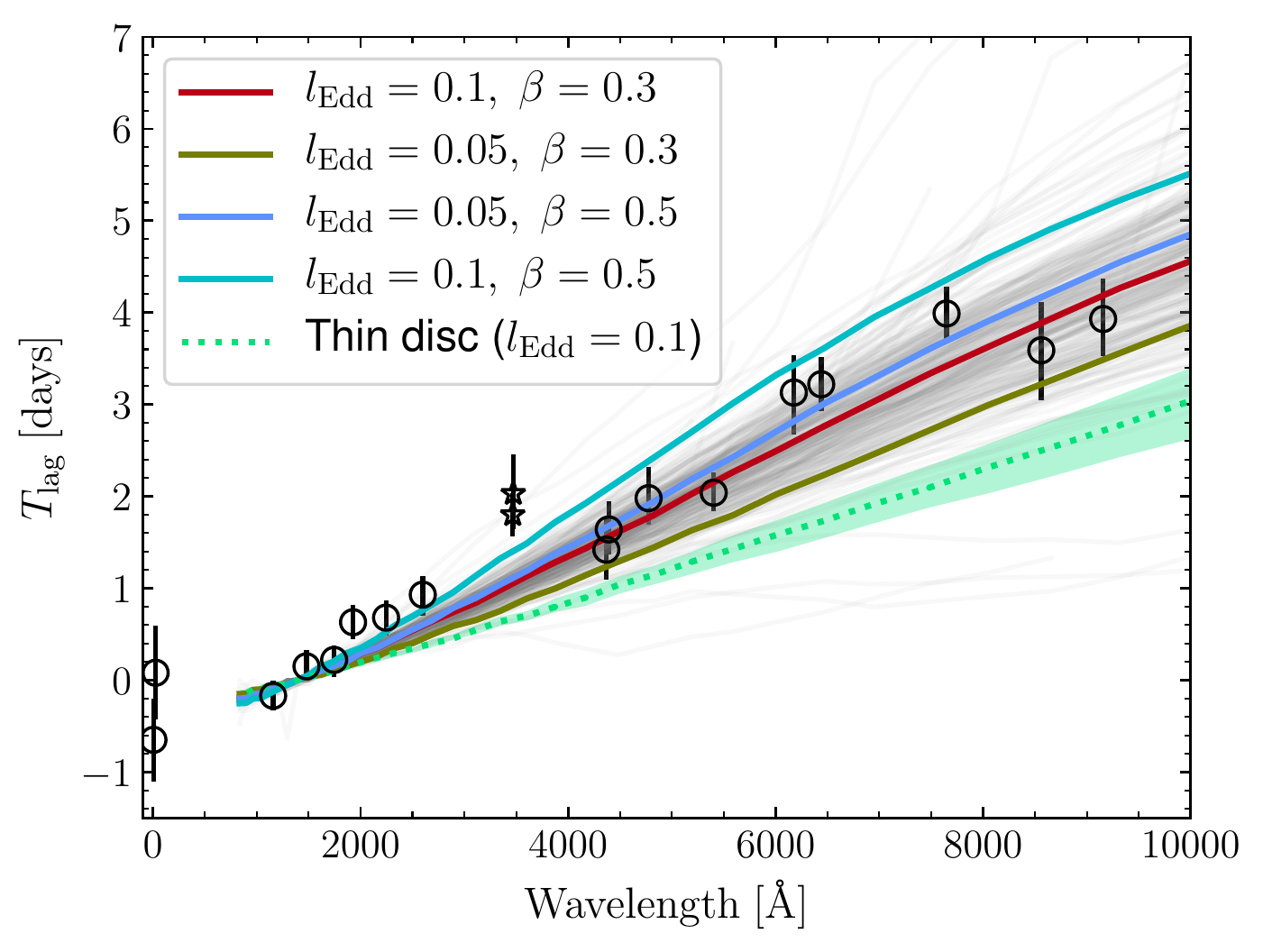}
\caption{Time lag (with respect to the $1367$ \AA\ continuum) as a function of rest-frame 
wavelength for NGC 5548 \citep{Edelson2015, Fausnaugh2016}. The two star symbols 
denote the time lags of $u$ and $U$ bands; they are outliers since the two bands are 
significantly contaminated by BEL emission \citep{Fausnaugh2016}. Each 
of the $256$ X-ray reprocessing simulations for the wind case with $\beta=0.3$ and 
$l_{\mathrm{Edd}}=0.1$ is shown as a gray solid curve. Due to reasons outlined in 
Section~\ref{sect:slope}, the time lag-wavelength relation varies per simulation. The 
red thick solid curve represents the median time lag-wavelength relation of the $256$ 
simulations. The remaining three thick solid curves indicate the median time lag-wavelength 
relations for other wind cases. For comparison, we also show the reprocessing of a 
classical thin disc with the same bolometric luminosity (the dotted curve; the shaded region 
indicates the $1\sigma$ uncertainties derived from the $256$ simulations).}
\label{fig:tlag}
\end{figure}

We fit the $1367$ \AA\ light curve with the \textit{continuous time first-order 
autoregressive process} (i.e., CAR(1), whose PSD has the following shape $\mathrm{PSD}(f) 
\propto 1/(f_0^2+f^2)$, where $f_0$ is a characteristic frequency) via 
\textit{CARMA}\footnote{This package can be downloaded from 
\url{https://github.com/brandonckelly/carma_pack}.} \citep{Kelly2014}. We then generate 
mock light curves with a cadence of $0.1$ day using \textit{CARMA}. The baseline of a mock 
light curve is $180$ days, which is similar to the observations of NGC 5548. The mock light 
curves heat the underlying accretion disc with time delays. The time lag is \citep{Starkey2017}
\begin{equation}
\label{eq:tau}
\tau = \frac{\sqrt{H^2 + R^2} + H\cos i - R\cos \theta \sin i}{c} \ ,
\end{equation}
where $\theta$ is the azimuthal angle of the accretion disc. The temperature profile of the accretion 
disc is calculated at each epoch of the mock light curves. We can then obtain the mock light curves 
of different wavelengths by simply integrating the blackbody radiation over the whole disc. To obtain 
the time lags relative to the $1367$ \AA\ light curve as a function of wavelength, we adopt 
ICCFs \citep[the interpolation cross-correlation function, which shows the correlation coefficient 
($\rho$) as a function of time delay; see, e.g.,][]{Peterson1998}.\footnote{We use PYCCF, Python 
Cross Correlation Function for reverberation mapping studies, to calculate the ICCFs. For details, 
see \url{http://ascl.net/code/v/1868}. } The time lags are estimated from the centroid of the 
ICCF, defined as the $\rho$-weighted lag for which $\rho>0.8\rho_{\mathrm{max}}$. For 
each wind case, we repeat the reprocessing simulation $256$ times. For comparison, we also 
apply the reprocessing simulation to the thin disc; this simulation is also repeated $256$ times. 

In Fig.~\ref{fig:tlag}, we show the time lag as a function of wavelength for our four cases. In near-IR 
bands, the relation is shallower because the outer boundary of the disc, which is fixed to $2000\ 
R_{\mathrm{S}}$, limits the time lags (i.e., Eq.~\ref{eq:x} over-predicts $X$).\footnote{It is unlikely 
that the outer boundary can be significantly larger than $2000\ R_{\mathrm{S}}$ since self gravity 
will truncate the accretion disc.} Due to reasons outlined in 
Section~\ref{sect:slope}, for fixed physical parameters (i.e., $\beta$, $l_{\mathrm{Edd}}$, 
$M_{\mathrm{BH}}$, $i$ and inner and outer boundaries), the time lag-wavelength relation varies 
per simulation (the gray curves represent the $256$ simulations for $\beta=0.3$ and $l_{\mathrm{Edd}} 
=0.1$). We then calculate the median time lag-wavelength relations (i.e., the thick solid curves). 
For $\beta=0.3$ and $l_{\mathrm{Edd}}=0.1$, the slope of the median relation is $1.12\pm 0.15$, 
which is shallower than $4/(3-0.3)=1.48$;  this case can fit the observed UV-optical-IR time lags 
reasonably well. Meanwhile, the case with $\beta=0.5$ and $l_{\mathrm{Edd}}=0.05$ can also explain the 
observed UV-optical-IR time lags. If we increase/decrease $\beta$ or $l_{\mathrm{Edd}}$, the inter-band time 
lags will be larger/smaller. The case with $\beta=0.5$ and $l_{\mathrm{Edd}}=0.1$ ($\beta=0.3$ and 
$l_{\mathrm{Edd}}=0.05$) over-(under-)predicts the UV-optical-IR time lags by an overall factor of 
$\sim 1.2$. 

\begin{figure}
\includegraphics[width=\columnwidth]{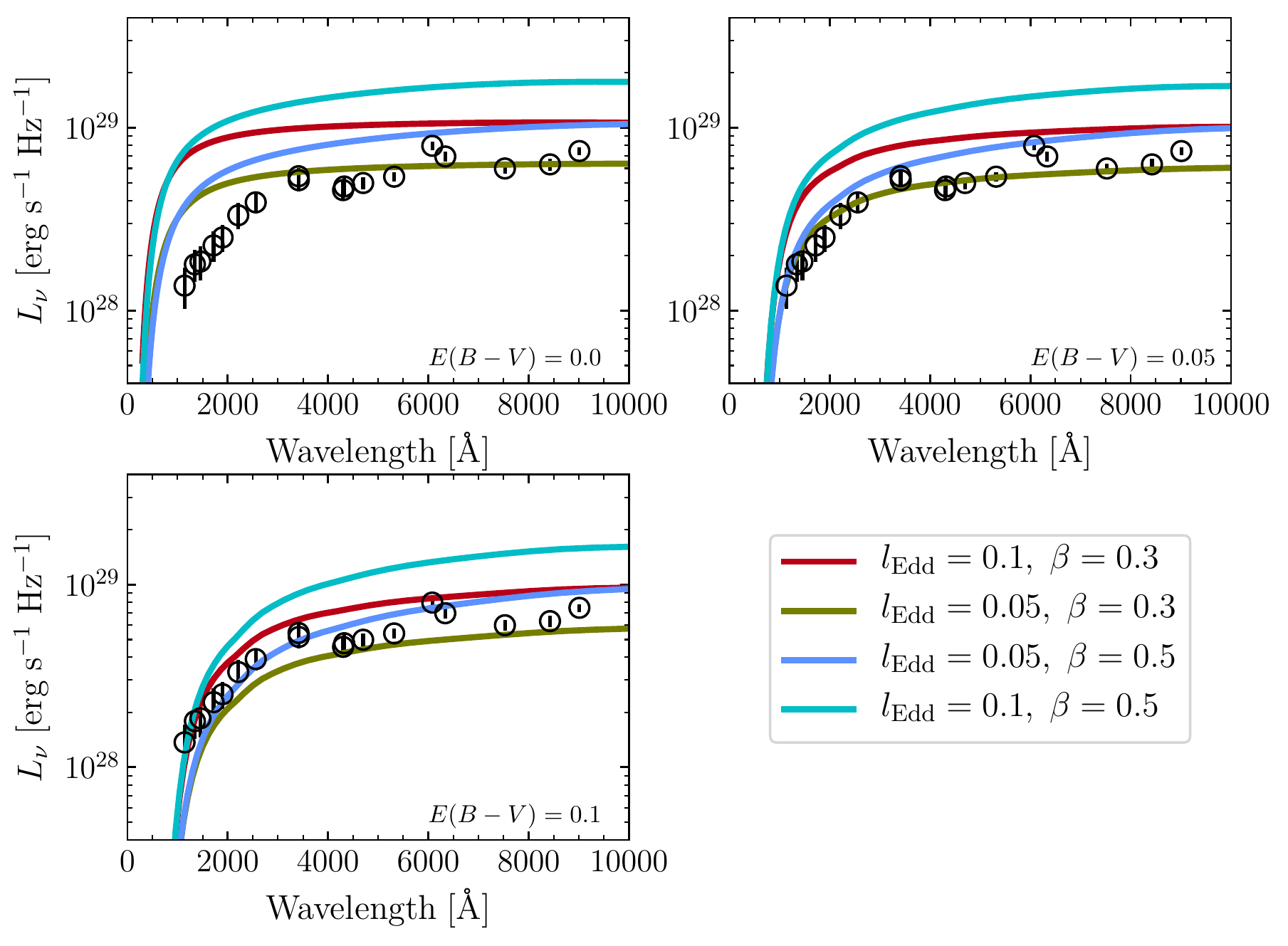}
\caption{Broadband SED of NGC 5548 from \protect\cite{Fausnaugh2016}, which is corrected 
for Galactic extinction with a Milky Way extinction law \protect\citep{Cardelli1989} and $E(B-V) 
=0.0171$ mag \protect\citep{Schlafly2011}. The solid curves indicate the SEDs 
derived from our disc-with-wind models. Without intrinsic extinction, these SEDs (the upper-left panel) 
generally over-predict the observations. However, such a discrepancy can be 
reduced if the SED has SMC-like intrinsic extinction. For $E(B-V)=0.05$ 
and $0.1$ mag, the resulting SEDs are presented in the upper-right and lower-left panels, 
respectively.}
\label{fig:sed}
\end{figure}

We can calculate the spectral energy distribution (SED) of a disc that suffers from winds by simply 
integrating the blackbody emission of the temperature profile of Eq.~\ref{eq:teff} over the whole 
accretion disc. In Fig.~\ref{fig:sed}, we plot the luminosity density $L_{\mathrm{\nu}}$ versus $\lambda$ 
from our models (the thick solid curves). The cases with $\beta=0.3$ and $l_{\mathrm{Edd}} 
=0.1$ or $\beta=0.5$ and $l_{\mathrm{Edd}}=0.05$ over-predict the optical-IR-UV emission by a 
factor of $2\sim 5$ (the upper-left panel of Fig.~\ref{fig:sed}). 

It has been demonstrated that the classical thin disc with multi-temperature blackbody 
emission cannot fit all AGNs; additional modifications should be applied, including the disc wind and 
intrinsic extinction \citep{Capellupo2015}. In this work, the disc wind has been considered. The 
intrinsic extinction of NGC 5548 is not reliably determined. However, there is evidence that the intrinsic 
extinction is significant. For instance, \cite{Wamsteker1990} argue that an extinction level of $E(B-V) = 
0.05$ mag exists. Meanwhile, the Balmer decrement of NGC 5548 \citep[see table 2 of][]{Bentz2010} 
also indicates significant intrinsic extinction \citep{Dong2008}. We consider 
the Small Magellanic Cloud \citep[SMC;][]{Gordon2003} extinction law with $E(B-V)=0.05$ mag 
(the upper-right panel of Fig.~\ref{fig:sed}) or $E(B-V)=0.1$mag (the lower-left panel of Fig.~\ref{fig:sed}). 
The SMC law alters the SEDs by preferentially reducing the UV emission; the resulting SEDs are more 
consistent with observations. For the cases with $\beta=0.3$ and $l_{\mathrm{Edd}} =0.1$ or $\beta=0.5$ 
and $l_{\mathrm{Edd}}=0.05$, the differences between the theoretical SEDs and the observations are less 
than $\sim 50\%$ if the SMC law with $E(B-V)=0.1$ mag is applied. 

The cases that explain the observations of NGC 5548 (i.e., $l_{\mathrm{Edd}} 
=0.1$ and $\beta=0.3$ or $l_{\mathrm{Edd}} =0.05$ and $\beta=0.5$) predict a mass outflowing 
rate of $\sim 0.7\ M_{\odot}\ \mathrm{yr^{-1}}$. Such a wind might also be responsible for replenishing 
the observed warm absorber \citep[$\sim 0.3\ M_{\odot}\ \mathrm{yr^{-1}}$;][]{Ebrero2016} 
or the long-lasting clumpy outflow \citep{Kaastra2014}.

\section{Discussion}
\label{sect:dis}
The observed larger-than-expected inter-band time lags indicate that some key ingredient is missing 
in the classical thin disc theory. We propose that the missing ingredient can be wind. If an AGN 
accretion disc suffers from strong winds, the temperature of the outer disc should be higher than the 
classical thin disc theory in order to produce the same bolometric luminosity, and meanwhile the 
temperature profile is also shallower (see Fig.~\ref{fig:teff}). For a given band, the outer disc can 
contribute more to the observed flux. Therefore, our model can explain the observed 
larger-than-expected disc sizes (see Figs.~\ref{fig:x} and \ref{fig:tlag}) as well as the observed 
UV-optical-IR SED (see Fig.~\ref{fig:sed}). 

\subsection{What if the wind column density is not small?}
\label{sect:rt}
In previous sections, we consider that winds change the observed SED and time lags by modifying 
the temperature profile. Winds might also alter the emission from the underlying disc via radiative 
transfer if the optical depth is not small. Such radiative transfer effects can alter 
the continuum emission \citep[e.g.,][]{Murray1995, Fukue2007, You2016} and leave absorption/emission 
features in X-ray, UV and optical bands \citep[e.g.,][]{Murray1995, Sim2008, Kusterer2014, 
Matthews2015, Matthews2016, Fukumura2017}. The results depend on many parameters, 
especially the column density and ionization level. Full calculation of such effects are only possible 
for numerical simulations and beyond the scope of this work. However, we argue that these effects 
cannot change our main conclusion due to the following reasons. 

We can estimate the column density of wind, which is 
\begin{equation}
\label{eq:nh}
\begin{split}
& N_{\mathrm{H}}\sim \frac{\dot{M}_{\mathrm{out}}}{4\pi RV_{\mathrm{R}}m_p}=
\frac{\dot{M}_0r^{\beta}}{4\pi m_p\sqrt{GM_{\mathrm{BH}}R}} \\&
\sim 5\times 10^{23}r^{-0.2} \left(\frac{1-\beta}{0.7}\right) \left(\frac{r}{r_\mathrm{in}}\right)^{\beta-0.3} 
\left(\frac{l_{\mathrm{Edd}}}{0.1}\right)\ \mathrm{cm^{-2}} \ ,
\end{split}
\end{equation}
where the outflowing velocity $V_{\mathrm{R}}$ is assumed to be close to the Keplerian velocity; 
$m_p$ is the proton mass. This is only a rough estimate since the exact value depends on the 
wind geometry and line of sight. Note that Eq.~\ref{eq:nh} is roughly consistent with the empirical 
scaling relations of X-ray winds \citep{Tombesi2013}. 
It is well known that accretion disc winds can easily be full ionized by X-ray emission unless a 
``shielding'' gas is presented. In the outer radii ($\gtrsim 300\ R_{S}$), a line-driving 
wind might form only if an ``failed'' wind shields the X-ray emission \citep[e.g.,][]{Murray1995, 
Proga2004, Higginbottom2014}. It is unclear that this ``shielding'' gas can also prevent inner 
(i.e., from ISCO to a few thousand $R_S$) winds from being over-ionized. If not, winds are 
close to being fully ionized. The Thomson optical depth is 
\begin{equation}
\label{eq:depth}
\tau_{\mathrm{T}}=\sigma_{\mathrm{T}}N_{\mathrm{H}}\sim 0.33 r^{-0.2} \left(\frac{1-\beta}{0.7}\right) 
\left(\frac{r}{r_\mathrm{in}}\right)^{\beta-0.3} \left(\frac{l_{\mathrm{Edd}}}{0.1}\right)\ \ ,
\end{equation}
where $\sigma_{\mathrm{T}}=6.65\times 10^{-25}\ \mathrm{cm}^2$ is the Thomson 
cross section. For NGC 5548, Eq.~\ref{eq:depth} indicates that the wind is transparent to 
UV-optical-IR photons. Meanwhile, the outflowing mass rate we assume should be regarded as an 
upper limit. This is because other mechanisms \citep[e.g.,][]{Hall2018} might also increase disc sizes. 
Variable hard X-ray or FUV photons could illuminate the underlying disc and induce longer-wavelength 
variations. 

Alternatively, if Eq.~\ref{eq:depth} under-estimates the optical depth (i.e., the wind 
is optically thick to disc emission), the photosphere is not the surface of the underlying disc but the wind; 
the effective temperature decreases. In some numerical simulations \citep[e.g.,][]{Proga2000, 
Fukumura2014}, winds are found to be equatorial with $H/R \lesssim 0.1$; their densities drop rapidly 
with increasing height. Therefore, the size of the photosphere might be only slightly larger than the 
surface of the disc and the effective temperature might decrease only by a negligible amount.

\subsection{Comparing our model with others}
The disc-with-wind scenario has been proposed before to estimate the AGN mass growth rate and 
SED \citep{Slone2012, Laor2014}. Winds are expected from theoretical arguments 
\citep[e.g.,][]{Blandford1982, Jiao2011, Cao2013, Gu2015}, numerical simulations \citep[e.g.,][]{Ohsuga2009, 
Yuan2012a, Yuan2012b, Yuan2015, Mou2017} and observational results \citep[e.g.,][]{Weymann1991, 
Murray1995, Trump2006, Richards2011, Tombesi2013, Filiz2014, Grier2015, Sun2018c}. 
However, it has not been demonstrated until this work that the sizes of emission regions 
of such a windy disc are larger than those of the classical thin disc \citep[but see also][]{Li2018}. 

In this work, we do not include a detailed discussion of the physics of winds. If our model is correct, 
we might use the observed disc sizes to infer wind strength $\beta$ and test wind models. For 
instance, in the model of winds accelerated by the magnetic fields, $\beta \leq 1/3$ \citep{Cao2013}. 
However, $\beta$ and $\dot{M}_0$ are degenerate. In this work, we fix $\beta=0.3$ to explain the 
observations of NGC 5548. In principle, we can also assume a larger value of $\beta$ (e.g., $\beta= 
0.5$) and a smaller $\dot{M}_0$ to match the inter-band time lags of NGC 5548. 

Our model can also be applied to other sources, such as NGC 4151 \citep{Edelson2017}, NGC 4593 
\citep{McHardy2018} or other studies \citep{Jiang2017, Mudd2017, Fausnaugh2018,  Homayouni2018, 
Kokubo2018}. It also has the potential to reconcile microlensing observations of quasars with the 
classical thin accretion disc theory, which is detailed by \cite{Li2018}.  

Our model is not the only one that can explain the observed time lag-wavelength relation of NGC 
5548. \cite{Cai2018} proposed a phenomenological model without X-ray reprocessing to explain 
the inter-band time lags. Instead of reprocessing, they attributed the time lags to 
the modulation of local temperature fluctuations by a speculative common temperature fluctuation, 
whose origin remains unclear. Meanwhile, \cite{Hall2018} suggested that, if the atmospheric 
density of an accretion disc is sufficiently low, scattering in the atmosphere can convert UV-optical 
photons into higher-energy ones and produce apparently larger disc sizes. However, their model 
failed to match the UV-X-ray SED of NGC 5548. They also speculated that this inconsistency can 
be resolved if disc suffers from winds and the emission from the innermost regions is suppressed.  

Our model cannot explain the large (i.e., a few days) time delay between hard X-ray and UV light 
curves. One possibility is that the observed time lag is contaminated by diffusion continuum emission 
from BEL region \citep{McHardy2018, Sun2018b}. It is also possible that the observed time delay 
does not correspond to the light-travel timescale but other timescales \citep[e.g., the dynamical or 
thermal timescale of a UV torus; see, e.g.,][]{Edelson2017, Gardner2017}.  

All in all, our model can help decipher the UV-optical-IR time lags of current continuum reverberation 
mapping results. Our model can also be tested by future continuum reverberation mapping of a large 
sample of AGNs. For instance, unlike the classical thin disc, our windy disc predicts a shallower 
$\tau$-$M_{\mathrm{BH}}$ relation and a steeper $\tau$-luminosity (or $\dot{M}$) relation (see 
Eq.~\ref{eq:rlam}). In addition, our model can be compared with reverberation mapping 
predictions from disc wind models \citep[e.g.,][]{Chiang1996, Waters2016, Mangham2017} and  
future observations of disc wind reverberation.

\section*{Acknowledgements}
We thank the anonymous referee for his/her helpful comments that improved the paper. 
We thank J. X. Wang, F. Yuan and Z. Y. Cai for valuable discussions. 
M.Y.S. and Y.Q.X. acknowledge the support from NSFC-11603022, NSFC-11473026, NSFC-11421303, 
the 973 Program (2015CB857004), the China Postdoctoral Science Foundation (2016M600485), the 
CAS Frontier Science Key Research Program (QYZDJ-SSW-SLH006). 

We made use of the following Python packages: Astropy \citep{Astropy2018}, CARMA \citep{Kelly2014}, 
Matplotlib \citep{Hunter2007}, Numpy \& Scipy \citep{scipy}, PYCCF \citep{Sun2018a}.

\bsp	
\label{lastpage}
\end{document}